\documentclass[floatfix, preprint, showpacs, showkeys, preprintnumbers, nofootinbib, superscriptaddress]{revtex4-1}
\usepackage{comment}
\usepackage{amssymb}
\usepackage{color}
\usepackage{rotating}
\usepackage{amsmath}
\usepackage{ulem}
\usepackage{overpic}
\usepackage{graphicx}
\usepackage{dcolumn}
\usepackage{bm}
\usepackage{epstopdf}
\usepackage{chngpage}
\usepackage{multirow}
\usepackage{slashed}
\usepackage{indentfirst}
\usepackage{booktabs}
\usepackage{amssymb,amsfonts}
\usepackage{amsmath}
\usepackage{color}
\usepackage[colorlinks,citecolor=blue,anchorcolor=blue,linkcolor=blue,hypertex,breaklinks=true]{hyperref}
\usepackage{CJKutf8}

\begin{document}
\begin{CJK}{UTF8}{<font>}
\title{Constraining nuclear symmetry energy with the charge radii of mirror-pair nuclei}
\author{Rong An}
\affiliation{Key Laboratory of Beam Technology of Ministry of Education, College of Nuclear Science and Technology, Beijing Normal University, Beijing 100875, China}
\affiliation{Key Laboratory of Beam Technology of Ministry of Education, Institute of Radiation Technology, Beijing Academy of Science and Technology, Beijing 100875, China}
\affiliation{CAS Key Laboratory of High Precision Nuclear Spectroscopy, Institute of Modern Physics, Chinese Academy of Sciences, Lanzhou 730000, China}
\author{Shuai Sun}
\affiliation{Key Laboratory of Beam Technology of Ministry of Education, College of Nuclear Science and Technology, Beijing Normal University, Beijing 100875, China}
\author{Li-Gang Cao}
\email[Corresponding author: ]{caolg@bnu.edu.cn}
\affiliation{Key Laboratory of Beam Technology of Ministry of Education, College of Nuclear Science and Technology, Beijing Normal University, Beijing 100875, China}
\affiliation{Key Laboratory of Beam Technology of Ministry of Education, Institute of Radiation Technology, Beijing Academy of Science and Technology, Beijing 100875, China}
\author{Feng-Shou Zhang}
\email[Corresponding author: ]{fszhang@bnu.edu.cn} 
\affiliation{Key Laboratory of Beam Technology of Ministry of Education, College of Nuclear Science and Technology, Beijing Normal University, Beijing 100875, China}
\affiliation{Key Laboratory of Beam Technology of Ministry of Education, Institute of Radiation Technology, Beijing Academy of Science and Technology, Beijing 100875, China}
\affiliation{Center of Theoretical Nuclear Physics, National Laboratory of Heavy Ion Accelerator of Lanzhou, Lanzhou 730000, China}

\begin{abstract}
The nuclear charge radius plays a vital role in determining the equation of state of isospin asymmetric nuclear matter.
  Based on the correlation between the differences in charge radii of mirror-partner nuclei and the slope parameter ($L$) of symmetry energy at the nuclear saturation density, an analysis of the calibrated slope parameter $L$ was performed in finite nuclei. In this study, relativistic and non-relativistic energy density functionals were employed to constrain the nuclear symmetry energy through the available databases of the mirror-pair nuclei $^{36}$Ca-$^{36}$S, $^{38}$Ca-$^{38}$Ar, and $^{54}$Ni-$^{54}$Fe. The deduced nuclear symmetry energy was located in the range 29.89-31.85 MeV, and $L$ of the symmetry energy essentially covered the range 22.50-51.55 MeV at the saturation density. Moreover, the extracted $L_s$ at the sensitivity density $\rho_{s}=0.10~\mathrm{fm}^{-3}$ was located in the interval range 30.52-39.76 MeV.
\end{abstract}


\maketitle
\section{Introduction}\label{section1}

Precise knowledge of nuclear symmetry energy (NSE), which is characterized as a component of the equation of state (EoS) of isospin asymmetric nuclear matter, can provide access to various physical phenomena relevant to a broad range of density profiles and energy scales~\cite{Steiner:2004fi,Li:2008gp}.
NSE plays an important role in understanding the nuclear structure. Moreover, the behavior of NSE may affect the properties of neutron stars~\cite{PhysRevC.100.045801,sunbaoyuan2018,Xu2022} and help to comprehend the supernova explosion mechanism and stellar nucleosynthesis in astrophysical studies~\cite{LATTIMER2007109}.

The density dependence of NSE, that is, $E_{\mathrm{s}}(\rho)$, can be expanded around the saturation density $\rho_{0}$ ($\simeq0.16$~fm$^{-3}$) as follows:
\begin{equation}
\label{eq:1}
E_{\mathrm{s}}(\rho)\approx{E_{\mathrm{s}}(\rho_{0})}+\frac{L}{3}\left(\frac{\rho-\rho_{0}}{\rho_{0}}\right)+\frac{K_{\mathrm{sym}}}{18}\left(\frac{\rho-\rho_{0}}{\rho_{0}}\right)^{2}+\dots,
\end{equation}
where $L$ and $K_{\mathrm{sym}}$ are the slope and curvature of the symmetry energy at the nuclear saturation density $\rho_{0}$, respectively.
The symmetry energy is believed to be associated with the isovector sensitive indicators in the EoS of isospin asymmetric systems.
Unfortunately, a direct connection between the experimental observables and the EoS is not possible.
Consequently, the microscopic implications of NSE can be extracted indirectly from the ground and the collective excited state properties of atomic nuclei, reaction observables, and detected dense astrophysical events~\cite{BALDO2016203,RocaMaza:2018ujj}.

Thus far, enormous efforts have been undertaken to determine the EoS over the spread of density profiles and energy scales~\cite{PhysRevC.86.015803}.
The neutron skin thickness (NST) of a heavy nucleus provides an available constraint on the EoS of neutron-rich matter around $\rho_{0}$~\cite{PhysRevLett.102.122502,PhysRevC.93.064303,PhysRevC.81.051303,ZHANG2013234,Liu_2011,scienceZuowei,PhysRevC.90.064310,xujun2020}.
In the laboratory, the radius of $^{208}$Pb has been detected by measuring the parity-violating asymmetry in the polarized elastic electron scattering experiments, for example sequentially in the 
PREX-II~\cite{PhysRevLett.126.172502}. The accuracy of NST has been further updated from the latest performance, namely ${R}_{\mathrm{skin}}^{208}=0.283\pm$0.071 fm.
The behavior of $E_{\mathrm{s}}(\rho)$ is mostly governed by the slope of the symmetry energy, $L$. 
The correlation between ${R}_{\mathrm{skin}}^{208}$ and the slope parameter $L$ leads to a value of $L=106\pm$37 MeV~\cite{PhysRevLett.126.172503}.
In addition, NSE can also be investigated using the isotope binding energy difference~\cite{DANIELEWICZ2003233,PhysRevC.82.064306} and double magic nuclei~\cite{PhysRevLett.111.232502,PhysRevC94.044322}.

Significant, progress has been made in evaluating NSE from the collective excited state properties of finite nuclei, such as isobaric analog states~\cite{DANIELEWICZ20141}, pygmy dipole resonance (PDR)~\cite{PhysRevC.81.041301}, electric dipole polarizability~\cite{PhysRevC.90.064317}, giant dipole resonance (GDR)~\cite{caocpl2008}, isovector giant quadrupole resonance (IVGQR)~\cite{PhysRevC.87.034301}, and charge-exchange giant resonance~\cite{PhysRevC.92.034308,PhysRevC.94.044313,KRASZNAHORKAY2013428,chengshihui}. The results for $L$ extracted from PDR in $^{68}$Ni and $^{132}$Sn were constrained to be in the intervals 50.3-89.4 MeV and $29.0$-$82.0$ MeV~\cite{PhysRevC.81.041301}, respectively. The deduced slope parameter from the weighted average can cover the range $L=$64.8$\pm$15.7 MeV. As suggested in Ref.~\cite{PhysRevC.87.034301}, the slope parameter of the symmetry energy can be reduced to 37$\pm$18 MeV by exploiting IVGQR energies.

Moreover, NSE offers a key requirement for our understanding of nuclear reactions under isospin diffusions and isotopic distributions~\cite{colonna2014,PhysRevLett.110.042701,Wei2022}.
Heavy-ion collisions (HICs) provide a sensitive probe to link the nuclear EoS which depends on isovector potentials.
In transport models, NSE is derived by simulating isospin-sensitive observables~\cite{wang2012,Akira2019,JHANG2021136016,PhysRevLett.126.162701,PhysRevLett.102.122701}. Hence, many simulation codes are desirable to determine the NSE~\cite{Feng:2009am,PhysRevLett.97.052701,tian2011,XIE20131510,XIE2014250,Yu2020}. More details about the transport simulations can be found in a recent study~\cite{PhysRevC.104.024603}.
Meanwhile, new observations of compact stellar objects have provided plentiful data that help discern the EoS across saturation densities~\cite{Miller_2019,jmakers2021}.
The range of $L$ can be deduced from the observation of dense object events, such as the production of gravitational waves from the neutron star merger GW170817~\cite{Raithel_2019}, resulting in $L=$11-65 MeV.

The density dependence of symmetry energy is fairly uncertain, except for the bulk properties at the saturation density $\rho_{0}$.
This challenges us in reducing the intrinsic uncertainties of the model from multiple aspects of the isovector components.
As demonstrated in Refs.~\cite{PhysRevC.88.011301,PhysRevLett.119.122502}, the difference in the root-mean-square (rms) charge radii of mirror-pair nuclei ($\Delta{R_{\mathrm{ch}}}$) obtained using Skyrme functionals provides an alternative opportunity for calibrating the density dependence of NSE.
\begin{table}[!htb]
\caption{$R_{\mathrm{ch}}$ and $\Delta{R_{\mathrm{ch}}}$ database for the $A = 36$, $38$, and $54$ mirror-pair nuclei. The parentheses beside the values of charge radii and the difference in charge radii are systematic uncertainties~\cite{Miller2019,PhysRevLett.127.182503,PhysRevResearch.2.022035}.}
\label{tab1}
\doublerulesep 0.1pt \tabcolsep 9pt
\begin{tabular}{cccc}
\hline
~~~~~$A$~~~~~ &~~~~~~ & ~~~~~$R_{\mathrm{ch}}$~(fm)~~~~~ & ~~~~~$\Delta{R_{\mathrm{ch}}}$~(fm)~~~~~ \\
\hline
  36  & Ca &  3.4484(27) &  \\
    & S  &  3.2982(12) & 0.150(4)\\
38  & Ca &  3.4652(17) & \\
    & Ar &  3.4022(15) & 0.063(3) \\
54  & Ni &  3.7370(30) & \\
    & Fe &  3.6880(17) & 0.049(4) \\
\hline
\end{tabular}
\end{table}
A related linear correlation between $\Delta{R_{\mathrm{ch}}}$ and the slope parameter $L$ has been established.
In Ref.~\cite{PhysRevResearch.2.022035}, the differences in the charge radii of the mirror-partner nuclei $^{36}$Ca-$^{36}$S and $^{38}$Ca-$^{38}$Ar were investigated with varied values of $L$.
It is evident that the slope parameter lies in the range $L=$5-70 MeV.
The latest precise determination evaluated the correlation between the difference in charge radii of the mirror-partner nuclei $^{54}$Ni-$^{54}$Fe and the slope parameter, which implied a range of $L=$21-88 MeV~\cite{PhysRevLett.127.182503}.
The rms charge radii of the nuclei $^{54}$Ni and $^{54}$Fe were obtained using Skyrme energy density functionals (EDFs) and  covariant density functional theories (CDFTs), respectively.

As demonstrated above, $\Delta{R_{\mathrm{ch}}}$ of mirror-pair nuclei can be employed to extract information about $L$.
The latest results of the charge radii of $^{54}$Ni can facilitate efficient exploration of the nuclear EoS.
The experimental data for $R_{\mathrm{ch}}$ and $\Delta{R_{\mathrm{ch}}}$ of the corresponding mirror partners are listed in Table~\ref{tab1}~\cite{Miller2019,PhysRevLett.127.182503,PhysRevResearch.2.022035}.
The NSE characterized as an isovector indicator in effective interactions should be systematically evaluated based on the latest experiments.
To further obtain a comprehensive conclusion about the correlations between $L$ and $\Delta{R_{\mathrm{ch}}}$, $\Delta{R_{\mathrm{ch}}}$ for the pairs of mass numbers $A=36$, $38$, and $54$ can be calculated using non-relativistic and relativistic (covariant) EDFs.
Although the correlations between the incompressibility coefficients and isovector parameters are generally weaker than the correlations between the slope parameter $L$ and symmetry energy~\cite{PhysRevC.104.054324}, the uncertainty suffered from nuclear incompressibility is inevitable in the evaluated process.
Therefore, the values of incompressibility coefficients characterized as isoscalar parameters are almost identical for the two types of EDFs.

The reminder of this paper is organized as follows: In Sec.~\ref{sec1}, we briefly report the theoretical models. In Sec.~\ref{sec2}, we present the results and discussion. Finally, a summary and an outlook are provided in Sec.~\ref{sec3}.

\section{Theoretical framework}\label{sec1}
In this study, we adopted two types of widely used nuclear density functionals to extract information about the nuclear matter EoS, namely, the sophisticated Skyrme and covariant EDFs.
Both have achieved great success in describing the structures of finite nuclei, especially global properties such as binding energies and charge radii. For a detailed introduction to non-relativistic and relativistic EDFs, refer to Refs.~\cite{Bender:2003jk,VRETENAR2005101,liang2015}. In this paper, we briefly introduce the two nuclear density functionals. The effective interaction in sophisticated Skyrme-type EDFs, which is expressed as an effective zero-range force between nucleons with density-dependence and momentum-dependence terms, is as follows:~\cite{CHABANAT1997710,Chabanat1998231}
\begin{eqnarray}
V(\mathbf{r}_{1},\mathbf{r}_{2})&=&t_{0}(1+x_{0}\mathbf{P}_{\sigma})\delta(\mathbf{r})\nonumber\\
&&+\frac{1}{2}t_{1}(1+x_{1}\mathbf{P}_{\sigma})\left[\mathbf{P}'^{2}\delta(\mathbf{r})+\delta(\mathbf{r})\mathbf{P}^{2}\right]\nonumber\\
&&+t_{2}(1+x_{2}\mathbf{P}_{\sigma})\mathbf{P}'\cdot\delta(\mathbf{r})\mathbf{P}\nonumber\\
&&+\frac{1}{6}t_{3}(1+x_{3}\mathbf{P}_{\sigma})[\rho(\mathbf{R})]^{\alpha}\delta(\mathbf{r})\nonumber\\
&&+\mathrm{i}W_{0}\mathbf{\sigma}\cdot\left[\mathbf{P}'\times\delta(\mathbf{r})\mathbf{P}\right],
\end{eqnarray}
where $\mathbf{r}=\mathbf{r}_{1}-\mathbf{r}_{2}$ and $\mathbf{R}=(\mathbf{r}_{1}+\mathbf{r}_{2})/2$ are related to the positions of two nucleons, $\mathbf{P}=(\nabla_{1}-\nabla_{2})/2\mathrm{i}$ is the relative momentum operator and $\mathbf{P'}$ is its complex conjugate acting on the left, $\sigma=\vec{\sigma}_{1}+\vec{\sigma}_{2}$, and $\mathbf{P_{\sigma}}=(1+\vec{\sigma}_{1}+\vec{\sigma}_{2})/2$ is the spin exchange operator.
The quantities $\alpha$, $t_{i}$, and $x_{i}$ ($i=0$-3) represent the parameters of the Skyrme forces.

For covariant EDFs, the interacting Lagrangian density has the following form:~\cite{PhysRevLett.86.5647,PhysRevC.64.062802}
\begin{eqnarray}
\mathcal{L}&=&\bar{\psi}[\mathrm{i}\gamma^\mu\partial_\mu-M-g_\sigma\sigma
-\gamma^\mu(g_\omega\omega_\mu+g_\rho\vec
{\tau}\cdotp\vec{\rho}_{\mu}+e\mathbf{A}_\mu)]\psi\nonumber\\
&&+\frac{1}{2}\partial^\mu\sigma\partial_\mu\sigma-\frac{1}{2}m_\sigma^2\sigma^2
-\frac{1}{3}g_{2}\sigma^{3}-\frac{1}{4}g_{3}\sigma^{4}\nonumber\\
&&-\frac{1}{4}\mathbf{\Omega}^{\mu\nu}\mathbf{\Omega}_{\mu\nu}+\frac{1}{2}m_{\omega}^2\omega_\mu\omega^\mu
+\frac{1}{4}c_{3}(\omega^{\mu}\omega_{\mu})^{2}\nonumber\\
&&-\frac{1}{4}\vec{R}_{\mu\nu}\cdotp\vec{R}^{\mu\nu}+\frac{1}{2}m_\rho^2\vec{\rho}^\mu\cdotp\vec{\rho}_\mu
\nonumber\\
&&+\Lambda_v(g_{\rho}^{2}\vec{\rho}_{\mu}\vec{\rho}^{\mu})(g_{\omega}^{2}\omega_\mu\omega^\mu)
-\frac{1}{4}\mathbf{F}^{\mu\nu}\mathbf{F}_{\mu\nu}.
\end{eqnarray}
Here, $\mathbf{\Omega}_{\mu\nu}=\partial_{\mu}\omega_{\nu}-\partial_{\nu}\omega_{\mu}$, $\vec{R}_{\mu\nu}=\partial_{\mu}\vec{\rho}_{\nu}-\partial_{\nu}\vec{\rho}_{\mu}-g_{\rho}(\vec{\rho}_{\mu}\times\vec{\rho}_{\nu})$, $\mathbf{F}_{\mu\nu}=\partial_{\mu}\mathbf{A}_{\nu}-\partial_{\nu}\mathbf{A}_{\mu}$, and $M$, $m_{\sigma}$, $m_{\omega}$, and $m_{\rho}$ are the nucleon, $\mathbf{\sigma}$, $\mathbf{\omega}$, and $\mathbf{\rho}$ meson masses, respectively. The quantities $g_{\sigma}$, $g_{\omega}$, $g_{\rho}$, $g_{2}$, $g_{3}$, $c_{3}$, and $e^{2}/4\pi=1/137$ are the coupling constants for the $\mathbf{\sigma}$, $\mathbf{\omega}$, $\mathbf{\rho}$ mesons, and photons, respectively. The parameter set $\Lambda_v$ represents the coupling strength between the $\omega$ and $\rho$ mesons.
Solving the Skyrme HF and Dirac equations on a spherical basis leads to the eigenenergies and wave functions of the constituent nucleons,from which the bulk properties of the ground states can be obtained using this standard procedure~\cite{Bender:2003jk,VRETENAR2005101,liang2015,CHABANAT1997710,Chabanat1998231,PhysRevLett.86.5647,PhysRevC.64.062802}.

\begin{figure}
\centering
\includegraphics[scale=0.4]{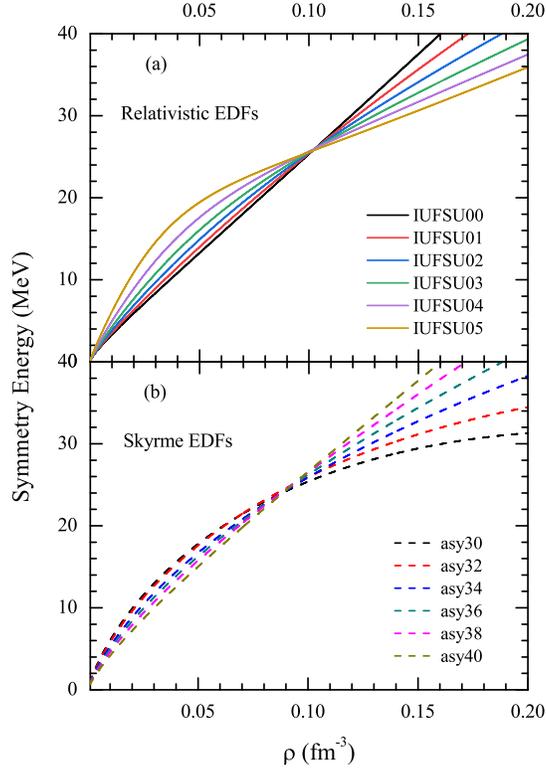}
\caption{(Color online)Symmetry energies characterized by relativistic EDFS (a) and Skyrme EDFs (b) plotted as a function of density. }
\label{fig21}
\end{figure}
In our applications, two families of parameter sets were adopted: the asy family for Skyrme EDFs~\cite{colo} and the IUFSU family for covariant EDFs~\cite{PhysRevC.83.034319}.
All of these effective forces were built by fitting the parameters to specific observables of finite nuclei, such as binding energies and charge radii, and the isovector part of the EoS was generated in such a way that the symmetry energy remained at a fixed value ($\approx$26 MeV) around a sensitivity density of $\rho\approx0.1$ fm$^{-3}$; thus, the interactions were characterized by different values of the symmetry energy at saturation density.
This procedure ensures that the quality of the fit cannot be contaminated and that all isoscalar observables remain unchanged, for example, the incompressibility coefficient almost equals to 230 MeV. In Fig.~\ref{fig21}, the density dependence behaviors of the symmetry energies are plotted using the relativistic and Skyrme EDFs. With increasing symmetry energy at the saturation density $\rho_{0}\simeq0.16~\mathrm{fm}^{-3}$, the slope became larger for these two families of parameter sets. For the symmetry energy around the sensitivity density $\rho_{s}=0.10~\mathrm{fm}^{-3}$, these values were almost unchanged. Further details can be found in Refs.~\cite{PhysRevC.83.034319,PhysRevC.70.024307}.
It is worth mentioning that both the symmetry energies and slope parameters can cover a wide range.
To reduce the intrinsic uncertainties of the models, such parameterization sets should be expected to provide stringent constraints on the observables that are highly sensitive to the density dependence of the symmetry energy.
The corresponding values of the bulk properties of the nuclear matter are shown explicitly in Table~\ref{tab2}.
\begin{table}[!htb]
\caption{Models used for the calculation of $R_{\mathrm{ch}}$. The corresponding bulk properties of nuclear matter used in this study, such as symmetry energy $E_{\mathrm{s}}$ (MeV), the slope parameter $L$ (MeV), and the incompressibility $K_{\mathrm{\infty}}$ (MeV) at saturation density, are given.}\label{tab2}
\doublerulesep 0.1pt \tabcolsep 5.8pt
\begin{tabular}{ccccc}
\hline
Type &   Sets & $E_{s}$~(MeV) & $L$~(MeV) & $K_{\mathrm{\infty}}$~(MeV)\\
\hline
RMF  &   IUFSU05 & 30.48 & ~46.11 & 229.98 \\
     &   IUFSU04 & 31.52 & ~52.09 & 229.98 \\
     &   IUFSU03 & 32.59 & ~60.52 & 230.05 \\
     &   IUFSU02 & 33.85 & ~71.83 & 230.01   \\
     &   IUFUS01 & 35.49 & ~87.27 & 230.04  \\
     &   IUFSU00 & 37.16 & 108.76 & 229.88 \\
\hline
Skyrme  &   asy30 &  30.00 &  ~22.87 & 230.20 \\
     &   asy32 &  31.99 &  ~36.22 & 229.99 \\
     &   asy34 &  33.99 &  ~56.14 & 229.84 \\
     &   asy36 &  36.00 &  ~71.54 & 229.93 \\
     &   asy38 &  38.00 &  ~87.62 & 230.20\\
     &   asy40 &  40.01 & 106.09  & 230.09\\
\hline
\end{tabular}
\end{table}

\section{Results and Discussions}\label{sec2}
\begin{figure}
\centering
\includegraphics[scale=0.4]{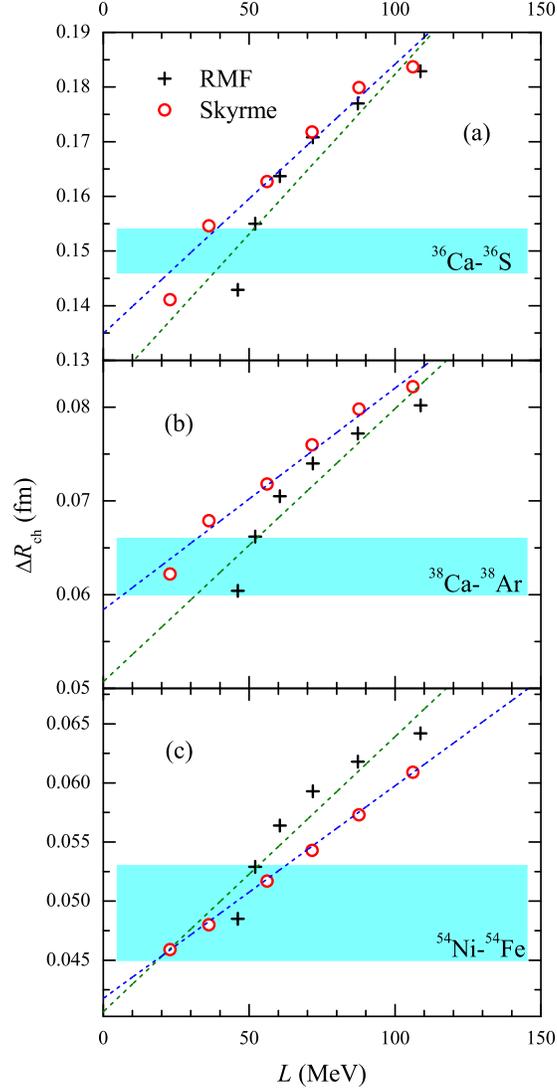}
\caption{(Color online) $\Delta{R}_{\mathrm{ch}}$ of the mirror-partner nuclei $^{36}$Ca-$^{36}$S (a), $^{38}$Ca-$^{38}$Ar (b), and $^{54}$Ni-$^{54}$Fe (c) as a function of the slope parameter $L$ at the saturation density $\rho_{0}$. The experimental result is shown as a horizontal light blue band. The crosses are results of relativistic EDFs, and the open circles are for the Skyrme EDF calculations. The dashed lines indicate theoretical linear fits.}
\label{fig1}
\end{figure}

The results for $\Delta{R_{\mathrm{ch}}}$ are plotted as functions of $L$ in Fig.~\ref{fig1}. The non-relativistic Skyrme approach and RMF theory were used to assess the correlation between $\Delta{R_{\mathrm{ch}}}$ and $L$, which are shown by the open circles and crosses in Fig.~\ref{fig1}, respectively. The horizontal light blue bands indicate the uncertainties of $\Delta{R_{\mathrm{ch}}}$, which correspond to 0.146-0.154~fm ($^{36}$Ca-$^{36}$S), 0.060-0.066~fm ($^{38}$Ca-$^{38}$Ar), and 0.045-0.053~fm ($^{54}$Ni-$^{54}$Fe). The linear fits in the correction for $\Delta{R}_{\mathrm{ch}}$ are indicated by the dashed lines.

The results obtained using the Skyrme and covariant EDFs revealed an approximate linear correlation between $\Delta{R_{\mathrm{ch}}}$ and $L$.
Constraints on $L$ were deduced by comparing the theoretical predictions with the experimental results in Fig.~\ref{fig1}.
Note that the results for $A=54$ provided the slope of symmetry energy $L$ relevant to the range 17.99-62.43 MeV, whereas those for $A=38$ and $A=36$ were in the intervals 6.83-52.49 and 22.50-51.55 MeV, respectively.
We can see that these values essentially cover the theoretical uncertainties in Refs.~\cite{PhysRevResearch.2.022035,PhysRevLett.127.182503}.

\begin{figure}
\centering
\includegraphics[scale=0.4]{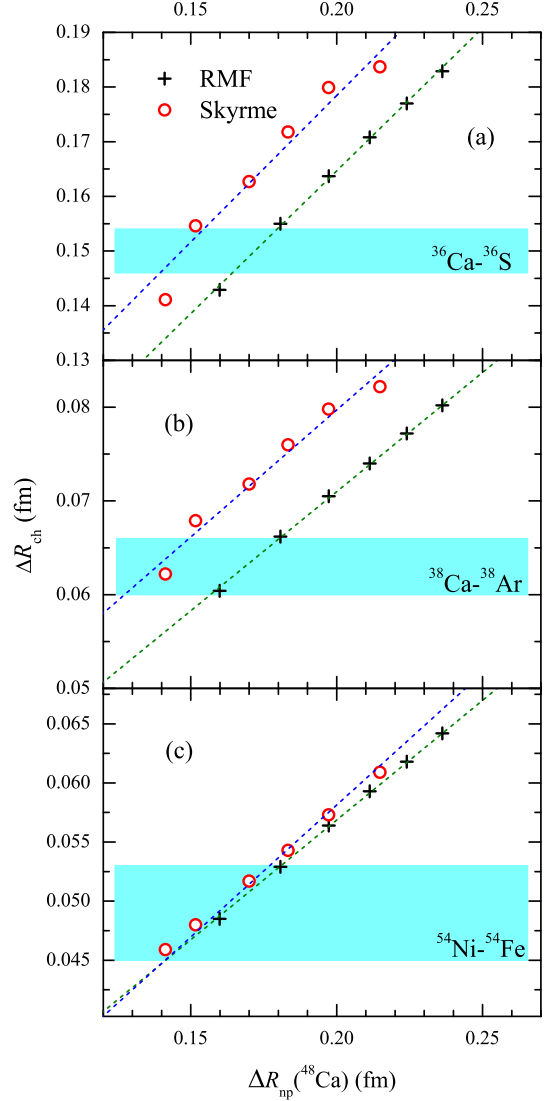}
\caption{(Color online) ``\ Data-to-data" relation between $\Delta{R_{\mathrm{ch}}}$ of the $A=36$, $38$, and $54$ mirror-partner nuclei and the neutron skin thickness $\Delta{R_{\mathrm{np}}}$ of $^{48}$Ca. The labels and color coding are the same as those used in Fig.~\ref{fig1}.}
\label{fig2}
\end{figure}
The NST of $^{48}$Ca is regarded as a feasible isovector indicator to constrain the EoS of nuclear matter.
The high-resolution $E1$ polarizability experiment performed at RCNP suggests that the NST of $^{48}$Ca is located in the interval $0.14\sim0.20$ fm~\cite{PhysRevLett.118.252501}.
Meanwhile, the CREX collaboration has reported a new value through parity-violating electron scattering measurements, namely, $0.071\sim0.171$ fm~\cite{PhysRevLett.129.042501}. The NST of $^{48}$Ca allows a direct comparison to microscopic
calculations using various slope parameters, $L$.
In Refs.~\cite{GAIDAROV2020122061,Sammarruca2018}, the linear correlation analysis of $\Delta{R}_{\mathrm{ch}}$ of mirror-partner nuclei and the corresponding NST has been clearly illustrated. Therefore, it is essential to evaluate the correlations between $\Delta{R}_{\mathrm{ch}}$ of mirror-pair nuclei and the NST of $^{48}$Ca.

The correlations between $\Delta{R_{\mathrm{ch}}}$ of the $A=36$, $38$, and $54$ mirror-partner nuclei and the NST of $^{48}$Ca ($\Delta{R_{\mathrm{np}}}$) are also shown in Fig.~\ref{fig2}.
The calculated $\Delta{R_{\mathrm{ch}}}$ substantially covered the current uncertainties of $\Delta{R_{\mathrm{np}}}$($^{48}$Ca) for both types of EDFs. For the $A=54$ mirror-pair nuclei, the Skyrme model gave a comparable correlation with respect to the RMF model.
A high linear correlation between $\Delta{R_{\mathrm{ch}}}$ of $A=54$ and $\Delta{R_{\mathrm{np}}}$ of $^{48}$Ca was found.
These correlations were also obtained using non-relativistic and relativistic EDFs for the $A=36$ and $38$ mirror partners, as shown in Figs.~\ref{fig2}~(a) and (b).
This means that information on symmetry energy can be extracted from the differences in the charge radii of mirror-pair nuclei.

\begin{figure}
\centering
\includegraphics[scale=0.45]{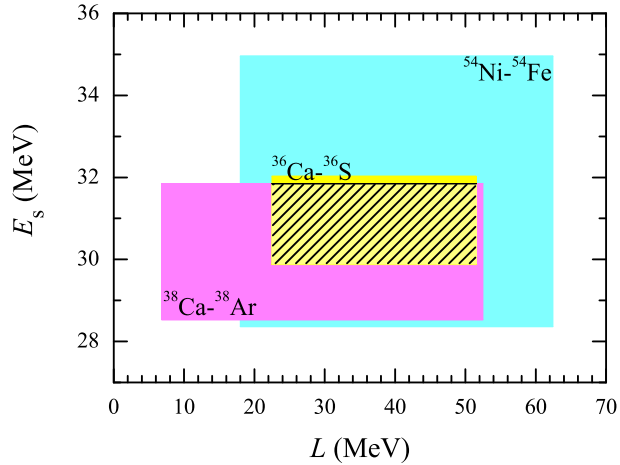}
\caption{(Color online) Symmetry energy $E_{\mathrm{s}}$ and the slope of symmetry energy $L$ are limited by $\Delta{R_{\mathrm{ch}}}$ of $A=36$ (yellow plane), $38$ (light purple panel), and $54$ (light blue plane) mirror-partner nuclei. The shadowed plane represents the result of the theoretical prediction in this study.}
\label{fig3}
\end{figure}

To obtain further constraints on the EoS of asymmetric nuclear matter, the relationship between the symmetry energy at saturation density and $\Delta{R_{\mathrm{ch}}}$ of mirror-pair nuclei can also be evaluated as well as the relationship between $L$ and $\Delta{R_{\mathrm{ch}}}$.
Therefore, the ``\ Data-to-data" relationships between the symmetry energies and the slope parameters are shown in Fig.~\ref{fig3} via various colored planes for the $A=36$ (Ca-S), $38$ (Ca-Ar), and  $54$ (Ni-Fe) mirror-partner nuclei.
The shadowed plane represents the result of the theoretical prediction.
From this figure, it is noticeable that the deduced symmetry energy was located in the interval $E_{s}=29.89$-$31.85$ MeV, and the slope of the symmetry energy covered the range $L=22.50$-$51.55$ MeV.
It should be mentioned that the present value provides a tighter constraint on $L$ than those in Refs.~\cite{PhysRevResearch.2.022035,PhysRevLett.127.182503}. Although the mirror-partner nuclei for the masses $A=36$, $38$, and $54$ were simultaneously considered in our evaluated procedure, the effects of deformation and model uncertainties were not incorporated in our calculations as in Refs.~\cite{PhysRevResearch.2.022035,PhysRevLett.127.182503}. In forthcoming investigations, we will carefully study the effect of deformation on the charge radius of finite nuclei and estimate the model uncertainties using a comprehensive set of modern density functionals, which may change the present results.

\begin{figure}
\centering
\includegraphics[scale=0.42]{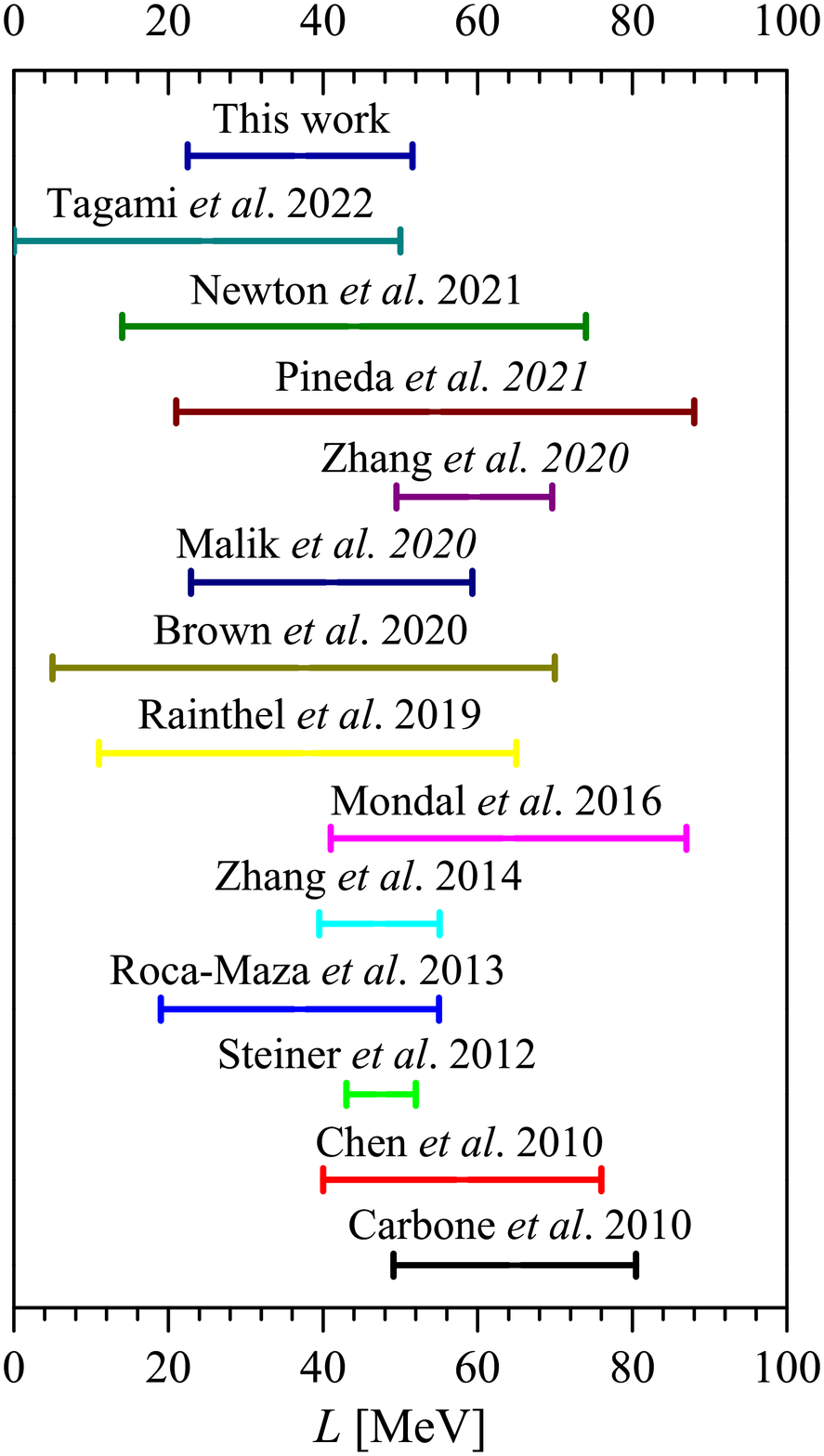}
\caption{(Color online) Comparison between the values of $L$ extracted in this study and those from existing literature. We partly compare the values extracted from various models: Carbone $et~al$.~\cite{PhysRevC.81.041301}, Chen $et~al$.~\cite{PRCChen024321}, Steiner $et~al$.~\cite{PhysRevLett.108.081102}, Roca-Maza $et~al$.~\cite{PhysRevC.87.034301}, Zhang $et~al$~\cite{PhysRevC.90.064317}, Mondal $et~al$.~\cite{PhysRevC.93.064303}, Raithel $et~al$.~\cite{Raithel_2019}, Brown $et~al$.~\cite{PhysRevResearch.2.022035}, Malik $et~al$.~\cite{PhysRevC.102.052801}, Zhang $et~al$.~\cite{PhysRevC101.034303}, Pineda $et~al$.~\cite{PhysRevLett.127.182503}, Newton $et~al$.~\cite{PhysRevC.103.064323}, and Tagami, $et~al$.~\cite{tagami106037}.}
\label{fig4}
\end{figure}
To facilitate the quantitative comparison of the extracted results with the theoretical calculations, various available estimates of the slope parameter $L$ of the symmetry energy are shown in Fig.~\ref{fig4}.
It is evident that our present result has a remarkable overlap with the results obtained by various methods or observables.
Our calculations predominantly covered the result for $L$ extracted from PDR in the $^{132}$Sn ($L=29.0$-$82.0$ MeV) but deviated from that of $^{68}$Ni ($L=50.3$-$89.4$ MeV)~\cite{PhysRevC.81.041301}.
Fig.~\ref{fig4} shows the weighted average value in the interval of $64.8\pm15.7$ MeV.
In addition, the electric dipole polarizability of a heavy nucleus is highly sensitive to both the magnitude and slope parameter of symmetry energy, providing a value of $L=47.3\pm$7.8 MeV~\cite{PhysRevC.90.064317}.
By exploiting this correlation together with the experimental values of the isoscalar and isovector giant quadrupole resonance (GQR) energies, the slope parameter of the symmetry energy was estimated as $L=37\pm$18 MeV~\cite{PhysRevC.87.034301}.
Both theoretical results essentially covered the current uncertainty in this study.

The NST $\Delta{r_{\mathrm{np}}}$ in heavy nuclei provides an alternative terrestrial probe to place restrictions on the nuclear EoS. In Ref.~\cite{PRCChen024321}, an accuracy value of $L=58\pm18$ MeV was deduced by analyzing the neutron skin data on Sn isotopes and the observables originating from HICs.
Our calibrated results covered the uncertainty of this value.
The latest NST of $^{48}$Ca detected by the CREX group yielded the slope parameter $L=0\sim50$ MeV~\cite{tagami106037}, which is in accordance with this study ($L=22.50$-$51.55$ MeV). Significantly, the NST of $^{208}$Pb detected by the PREX-II group yielded a larger value of $L=76\sim165$ MeV.
Reed et al.~\cite{PhysRevLett.126.172503} reported a comparable interval of the slope parameter $L=106\pm37$ MeV.
These values partially cover the interval range of $L=54\sim97$ MeV induced by HICs~\cite{PRC103.014616}. However, the evaluated range in this literature has no overlap with the slope parameter $L$ obtained by PREX-II.

In Ref.~\cite{PhysRevC101.034303}, more information, such as heavy-ion collision data, the neutron skin of $^{208}$Pb, tidal deformability, and the maximum mass of neutron dense objects, was used to calibrate the values of symmetry energy. This led to a symmetry energy slope of $L=59.57\pm10.06$ MeV, and the quantitative uncertainty was further reduced with respect to Ref.~\cite{PRCChen024321}.
Moreover, the slope parameter $L$ can be constrained by various observed astrophysical messages, except in the terrestrial nuclear experiments.
The currently available neutron star mass and radius measurements provide an important constraint on the EoS of neutron matter through quantum Monte Carlo simulations, in which the slope parameter $L$ is located in the range $43$-$52$ MeV~\cite{PhysRevLett.108.081102}.
The correlation of the tidal deformability of a neutron star with $L$ was studied using the sk$\Lambda$267 model, which gave the range $L= 41.1\pm$18.2 MeV~\cite{PhysRevC.102.052801}.
Moreover, the range $L= 11$-65 MeV was extracted from the observation of a gravitational-wave event of the neutron star merger GW170817~\cite{Raithel_2019}.
Our calibrated range ($L=22.50$-$51.55$ MeV) overlaps significantly with these deduced values.

Recently, emerging Bayesian frameworks have been developed widely to study the bulk properties of finite nuclei, for example, predictions in the nuclear charge radii~\cite{Dong:2021aqg,Dong:2022plb} and nuclear EoSs~\cite{xie2022}. The existing database of neutron skin and the bulk properties of nuclear matter are characterized by prior input quantities, which leads to credible values of $L=40^{+34}_{-26}$ MeV and $L= 37^{+9}_{-8}$ MeV~\cite{PhysRevC.103.064323}, respectively. All the evaluated values are consistent with the range of $L$ obtained in this study.

\begin{figure}
\centering
\includegraphics[scale=0.4]{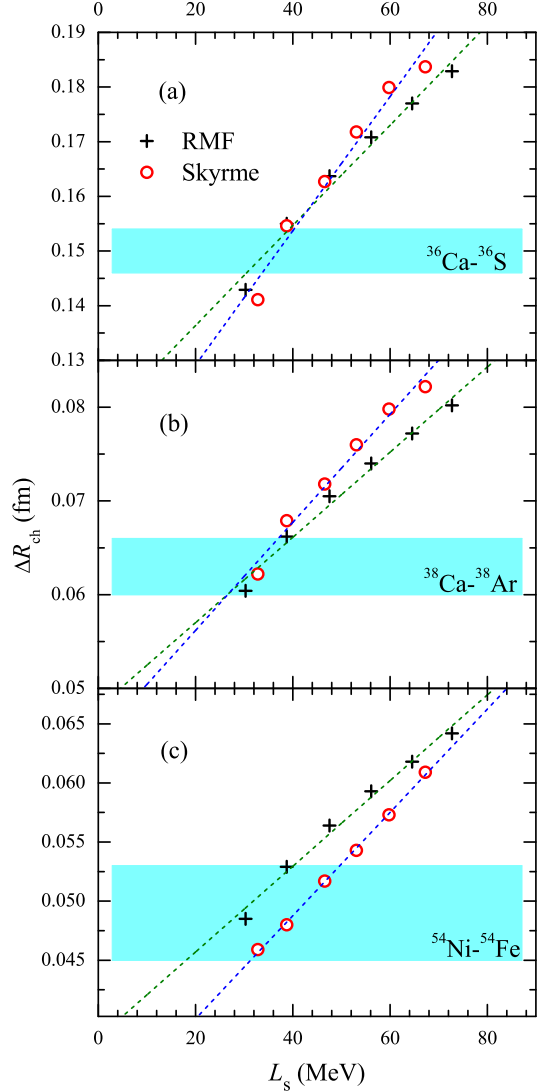}
\caption{(Color online) Same as Fig.~\ref{fig1}, but for the slope parameter $L_{s}$ at the sensitivity density $\rho_{s}=0.10~\mathrm{fm}^{-3}$.}
\label{fig31}
\end{figure}
As mentioned in Refs.~\cite{PhysRevC.89.017305,PhysRevC.84.044620}, the density dependence of the symmetry energy at the subsaturation density is associated with nuclear mass differences and multifragmentation production.
Thus, it is also interesting to give the constraint of the slope parameter $L_s$ at the sensitivity density $\rho_{s}=0.10~\mathrm{fm}^{-3}$ probed by the differences in the charge radii of mirror-partner nuclei ($\Delta{R}_{\mathrm{ch}}$).
In Fig.~\ref{fig31}, $\Delta{R}_{\mathrm{ch}}$ of the mirror-partner nuclei $^{36}$Ca-$^{36}$S (a), $^{38}$Ca-$^{38}$Ar (b), and $^{54}$Ni-$^{54}$Fe (c) as a function of the slope parameter $L_{s}$ at the sensitivity density $\rho_{s}=0.10~\mathrm{fm}^{-3}$ are plotted.
The highly linear correlations between $\Delta{R}_{\mathrm{ch}}$ and $L_s$ are also presented.
The extracted interval range of the slope parameter $L_{s}$ at the sensitivity density $\rho_{s}=0.10~\mathrm{fm}^{-3}$ was approximately 30.52¨C39.76 MeV.
This restricted value is relatively narrow compared to the interval range at the saturation density $\rho_{0}\simeq0.16~\mathrm{fm}^{-3}$.
\section{Summary and outlook}\label{sec3}
Microscopic methods based on families of non-relativistic and relativistic EDFs were employed to characterize a systematic variation of the isoscalar and isovector properties of corresponding nuclear matter EoSs. Our systematic analysis of the extraction of the slope parameter $L$ from the differences in mirror-partner nuclei charge radii provided a new result.
The slope parameter $L$ covered the interval range 22.50-51.55 MeV at the saturation density. Moreover, the slope parameter $L_s$ at the sensitivity density $\rho_{s}=0.10~\mathrm{fm}^{-3}$ lay in the interval range 30.52-39.76 MeV.
This led to us determining the density dependence of symmetry energy relatively accurately, which is a fundamental quantity for nuclear physics and for the implications in the study of neutron stars.

Linear fits were performed between the differences in the charge radii of mirror-partner nuclei and the slope parameter $L$. As suggested in Refs.~\cite{An2022,PhysRevC.105.014325,PhysRevC.101.021301,PhysRevC.103.054310,PhysRevC.102.024307}, precise descriptions of the nuclear charge radii are influenced by various mechanisms. Meanwhile, the precise measurement of the charge density distributions usually affects the NST of finite nuclei on a quantitative level. In Ref.~\cite{PhysRevC.105.L021301}, it is demonstrated that the differences in the charge radii of mirror-pair nuclei are systematically influenced by the pairing correlations. For weakly bound nuclei, configuration mixing should be discreetly considered when tacking pairing correlations~\cite{An_2018}. Hence, this study should be further reviewed.

\section{Acknowledgements}
This study was partly supported by the Key Laboratory of High Precision Nuclear Spectroscopy, Institute of Modern Physics, Chinese Academy of Sciences. This study was also supported in part by the National Natural Science Foundation of China under Grants No. 12135004, No. 11635003, and No. 11961141004. L.-G. C. acknowledges the support of the National Natural Science Foundation of China under Grants No. 12275025, and No. 11975096 and the Fundamental Research Funds for the Central Universities (2020NTST06).



\end{CJK}
\end{document}